%% file: main.tex
\documentclass[copyright, creativecommons]{eptcs}
\providecommand\event{GASCOM 2024}

\usepackage{bm}
\usepackage{underscore}  % From EPTCS' example file, for pdflatex
\usepackage[T1]{fontenc}  % From EPTCS' example file, for pdflatex

\input{preamble}

\title{Random Generation of Git Graphs}
\author{%
  Julien Courtiel%
  \institute{Normandie Univ, UNICAEN, ENSICAEN, CNRS, GREYC, 14000 Caen, France}%
  \email{julien.courtiel@unicaen.fr}%
  \and%
  Martin Pépin%
  \institute{Normandie Univ, UNICAEN, ENSICAEN, CNRS, GREYC, 14000 Caen, France}%
  \email{martin.pepin@unicaen.fr}%
}

\def\titlerunning{Random Generation of Git Graphs}
\def\authorrunning{J. Courtiel \& M. Pépin}
\begin{document}
\maketitle{}

\begin{abstract}
  \noindent
  Version Control Systems, such as
  Git and Mercurial, manage the history of a project as a Directed
  Acyclic Graph encoding the various divergences and synchronizations happening
  in its life cycle.
  A popular workflow in the industry, called the \textit{feature branch
  workflow}, constrains these graphs to be of a particular shape:
   a unique main branch, and non-interfering feature branches.

  Here we focus on the uniform random generation of those graphs with $n$ vertices, including $k$ on the main branch,  
  for which we provide three algorithms, for three different use-cases.
  The first, based on rejection, is efficient when aiming for small values of $k$ (more precisely whenever $k = O(\sqrt n)$). 
  The second takes as input any number $k$ of commits in the main branch, but requires costly precalculation.
  The last one is a Boltzmann generator and enables us to generate very large graphs while targeting a constant $k/n$ ratio.
  All these algorithms are linear in the size of their outputs.
\end{abstract}

\section{Motivation}

In software development, Version Control Systems (VCS in short) such as Git or Mercurial are crucial. 
They facilitate collaborative work by allowing multiple developers to concurrently contribute to a shared file system.
 VCS automatically save all project versions over time, along with the associated changes.

Most VCS offer \textit{branching} support, 
allowing developers to diverge from the main line of development and continue their work independently without affecting the main project line. 
These branches can be subsequently \textit{merged}, in order to integrate changes from one branch into another, like new features or bug fixes.

In the abstract, the history of a VCS repository can be seen as a Directed Acyclic Graph (DAG), where vertices are the different versions of the project (also named \textit{commits})
and arcs symbolize the changes between two versions. There are no restrictions on the shape of the graphs you can generate with a VCS, but
many projects follow a \textit{workflow},
that is a process and a set of conventions that define how branches are created, and how changes are integrated into the main codebase.

The purpose of this paper is to develop an efficient random sampler for DAGs that respect a parti\-cular workflow. 

One benefit of such a sampler would be to integrate  \textit{property-based tests} into VCS development. In these tests, instead of specifying explicit input values and expected outcomes, we define properties that should be satisfied for a wide range of repositories, which are generated randomly during the test. 
By generating diverse graph structures that adhere to the workflow's specifications, we ensure a comprehensive examination of the VCS's behavior according to plausible scenarios.
To give a concrete example based on work by one of the authors~\cite{CDL2023}, a random DAG sampler could experimentally check the effectiveness of \texttt{git bisect}, an algorithm that finds the commit where a bug has been introduced. 

In this paper, we will take a look at one of the simplest workflows, but one
that is widely used in the corporate world: the \emph{feature branch workflow}.
In this workflow, the non-main branches do not interfere with each other, and are
simply attached to the (unique) main branch. Here is a more formal definition of
graphs induced by this workflow. (This definition originally comes from~\cite{LecoqThesis}.)

\begin{definition}[\gitgraph]
  A \emph{\fullnamegitgraph} (or just \emph{\gitgraph}) is a DAG that consists of:
  \begin{itemize}
    \item a \emph{main branch}, that is a directed path of black vertices.
    \item potentially several \emph{feature branches}, that are directed paths that start and end on vertices of the main branch. The set of intermediary vertices is not empty and consists of white vertices. A black vertex cannot be the end point of several feature branches, just one at most (but it can be the starting point of several branches).
  \end{itemize}
\end{definition}

\begin{figure}[htb]
  \begin{center}
    \includegraphics[width=\textwidth]{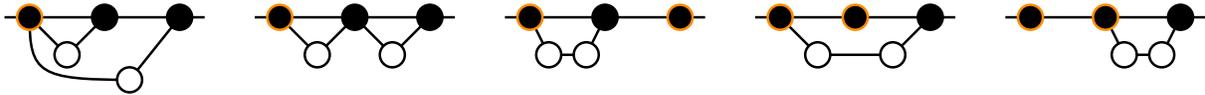}
    \caption{All \gitgraph{}s with $5$ vertices including $3$ black vertices. Edges are oriented from left to right. Free vertices are  outlined in orange.}
  \end{center}%
  \label{fig:allsize5}
\end{figure}

The size of a \gitgraph{}~$\gamma$ is its number of vertices. 
By convention, we assume that there exists a unique \gitgraph{} of size $0$.
Another important parameter is its number of black vertices, and will be denoted
by $\main \gamma$.
A black vertex is said to be \emph{free} if there is no feature branch ending on
it, \textit{i.e.}\ its indegree is at most $1$.
All \gitgraph{}s $\gamma$ of size $5$ with $\main \gamma = 3$ are listed in Figure~\ref{fig:allsize5}.

The fact that we forbid merges of multiple feature branches into the main one is
not a restriction of the VCS, but is advisable to maintain a clearer and more
understandable project history, reduce the risk of conflicts, and enhance
traceability and maintainability. This restriction is also discussed in~\cite{CDL2023}.

\section{The uniform model}

\subsection{A recursive decomposition}

We first describe a recursive decomposition of \gitgraph{}s, based on the number
of black vertices.
Consider the last black vertex $v_k$ of a \gitgraph\ of size $n$ and with $k>1$
black vertices. There are only two possibilities: either $v_k$ is free, or $v_k$
is a merge between the main branch and a feature branch (which is unique, by
definition).
In the latter case, the feature branch starts with a black vertex, which can be
any vertex of the main branch, but $v_k$.
Removing $v_k$ and the potential feature branch attached to it leads to a
smaller \gitgraph{}.

\begin{figure}[h!]
  \begin{center}
    \includegraphics[scale = 1]{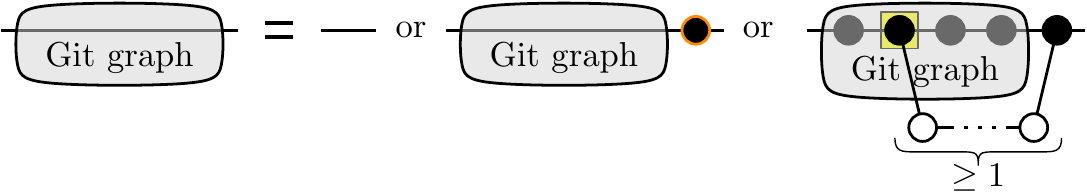}
    \vspace{-.5cm}
    \caption{How to decompose a \gitgraph.}%
    \label{fig:decomposition}
  \end{center}%
\end{figure}

By this reasoning, illustrated by Figure~\ref{fig:decomposition}, we obtain the
induction formula
\begin{equation}
   g_{n,k} = g_{n-1,k-1} + \sum_{\ell \geq 1} (k-1) g_{n-1-\ell,k-1}, 
   \quad \text{with }g_{0,k} =  \begin{cases} 1 & \text{if }k = 0 \\ 0 & \textrm{otherwise} \end{cases}
   \label{eq:rec:gnk}
\end{equation}
where~$g_{n,k}$ is the number of \gitgraph{}s with~$n$ vertices,~$k$ of them
being black.

This induction is sufficient to write a recursive generator (see~\cite{NW1978}
for the general theory of recursive samplers and~\cite{FZV1994} for a more
modern point of view in the context of the symbolic method).
We do not extend on this generator as we will present with Algorithm~\ref{algo:free} a more efficient sampler\footnote{
  The straightforward recursive algorithm obtained from this recurrence runs in~$O(nk)$ time, and requires the precomputation of~$O(n k)$ numbers of size~$O(n \log n)$, involving~$O(n^2 k)$ operations on big integers in total.
}, also based on the recursive method.

It is straightforward (especially if you are familiar with the symbolic
method~\cite{FS2009}) to translate Formula~\eqref{eq:rec:gnk} into a
differential equation whose solution is the generating function~\(G(z,u)\) of
\gitgraph{}s:
\begin{equation}
  G(z,u) = 1 + zu G(z,u) + \frac{z^2u^2}{1-z}\frac{\partial G}{\partial u}(z,u),
  \quad\text{where~}
  G(z, u) \coloneqq \sum_{n \geq 0}\sum_{k \geq 0} g_{n,k} z^n u^k.
  \label{eq:diffeq}
\end{equation}
Note that $G(z,u)$ is not analytic at $z=0$ since the number of \gitgraph{}s
grows as a factorial (we have~$g_{2k-1,k} \geq (k-1){!}$ by considering a
\gitgraph{} with only merge commits and feature branches with~$1$ white node).
For this reason, the previous equation does not seem to be usable for Boltzmann
sampling.

\subsection{Most \gitgraph{}s look alike under the uniform distribution}%
\label{sec:allalike}

A large random \gitgraph{} is with high probability of the same shape: 
about half of the commits are on the main branch, 
and most commits on the main branch are merges of size-$1$ branches.

\begin{proposition}
  Let~$u$ be any real positive number.
  Consider~$\gamma_n$ a random \gitgraph{} of size~$n$ taken with probabi\-lity
  $\dfrac{u^{\main{\gamma_n}}}{\sum_{\gamma \text{ \gitgraph\ of size }n}
  u^{\main \gamma}}$.
  Then the random variable $\frac{\main {\gamma_n}}{n}$ converges in probability to
  $\frac 1 2$ when $n$ goes to $+\infty$.
  (Note that~$u=1$ corresponds to the uniform distribution).
\end{proposition}

The intuition behind this result is that a large number of branches greatly increases the number of ways of connecting them to the main branch, hence favoring graphs with many short branches over ones with fewer but longer branches.

In particular, for any value of~$u$, the average number of commits in the main
branch is asymptotically equivalent to $\frac n 2$.
This motivates the introduction of a variant of this model which we detail in
the next half of this paper, and which allows more control over the number of
commits on the main branch.

\subsection{A rejection algorithm}\label{sec:rejection}

Before delving into the next model, it is worth noting that there is an
efficient rejection-based sampling algorithm for the case where~$k$ is small
based on the following inclusion.
Consider a variation~$\mathcal H$ of the model where 
every black vertex but the first one is the endpoint of a feature branch but 
it is allowed to have zero
commit on a feature branch. 
Denote by~$h_{n,k}$ the number of such graphs with~$n$ vertices including~$k$ on
the main branch.
Then \gitgraph{}s can be seen as a subset of these graphs by identifying empty
feature branches pointing at the first commit in~$\mathcal H$ with free commits
in \gitgraph{}s.
Moreover, whenever~$k \leq t\sqrt{n}$, for some constant~$t > 0$, we have that
\begin{equation*}
  c_t \le \frac{g_{n,k}}{h_{n,k}} \le 1
  \quad \text{for some~$c_t > 0$ that depends only on~$t$}.
\end{equation*}
The constant~$c_t$ is obtained by considering the class of \gitgraph{}s with only one free commit as a subset of \gitgraph{}s.
Explicit formulas exist for the cardinality of this class and for~$h_{n,k}$ and their ratio is~$\Theta(1)$ in the regime~$k \le t \sqrt n$.
This yields Algorithm~\ref{algo:rejection} for sampling uniform \gitgraph{}s
with a small main branch.
This algorithm can be implemented so as to perform~$O(k)$ array accesses
and~$O(k)$ RNG calls\footnote{%
  In practice, considering an RNG call to be~$O(1)$ faithfully reflects the
  runtime performance of such an algorithm.
  It is thus a realistic complexity model, that we use in the rest of this
  document.
  It is however important to note that every RNG call needs to produce
  about~$\log_2(n)$ random bits here.
}
in average.

\begin{algorithm}[h!]
  \caption{Rejection algorithm for \gitgraph{}s with $n$ vertices, $k$ of them being black}\label{algo:rejection}
  \begin{algorithmic}[1]
    \State{start with a chain of~$k$ black vertices}
    \State{arrange uniformly at random~$(n-k)$ white vertices into~$(k-1)$
      possibly empty chains}
    \State{attach the ends of these chains to the~$(k-1)$ last black vertices}
    \State{attach the start of every chain to a previous black vertex, chosen
    uniformly at random}
    \State{if any of the empty chains is not attached to the root, start over,
    otherwise return}
  \end{algorithmic}
\end{algorithm}

\section{The \labelled\ distribution}

\subsection{Description of the model}

Given the disadvantages of the uniform distribution, we propose a new model for random \gitgraph{}s that is easier to sample, gives more varied shapes, and with fine control over the number of black vertices.
The principle is that a \gitgraph\ $\gamma$ will have a probability to be generated proportional to \(u^{\main \gamma}/{\main \gamma}{!}\) where $u$ is a real positive parameter.

More precisely, we set \(\widetilde{G_n}(u) := \sum_{k = 1}^{n} g_{n,k} \frac{u^k}{k!}\) and \(\widetilde G (z,u) := \sum_{n \geq 0} \widetilde{G_n}(u) z^n\). Thus \(\widetilde G\) resembles an exponential generating function, but with a scaling of $k{!}$ instead of a scaling of $n{!}$. Unlike $G$ defined in the previous section, the function \(\widetilde G\) is analytic at $z=0$ (a direct consequence of Theorem~\ref{theo:tildeg} below).

\begin{definition}
  The probability under the \emph{\labelled\ distribution} of a \gitgraph\ of size $n$ and with $k$ black vertices is defined as \( \frac{u^k  z^n }{k! \widetilde G(z,u)}\), where $z$ and $u$ are positive parameters inside the domain of convergence of~$\widetilde G$.
\end{definition}

This is a multivariate Boltzmann model (exponential in~$u$ and ordinary in~$z$). A sampler based on this distribution falls into the category of \emph{Boltzmann generators}, for which a large number of results have been established, facilitating the generation of large objects~\cite{DFLS2004}.

By using the Borel transform~\cite{Borel1899} on Equation~\eqref{eq:diffeq}
with respect to the variable~$u$, that is to say the mapping~\( \sum_{n,k \geq 0} a_{n,k}
z^n u^k \mapsto \sum_{n,k \geq 0} \frac{a_{n,k} z^n u^k}{k!} \), we can obtain a
differential equation for~\(\widetilde G (z,u)\):
 \begin{equation}\label{eq:diffeqtildeg}
  \frac {\partial \widetilde G}{\partial u}(z,u) = z \widetilde G (z,u) + \frac{z^2 u}{1 - z} \frac{\partial \widetilde G}{\partial u}(z,u)
  \qquad
  \text{and} \qquad \widetilde G(z, 0) = 1.
\end{equation}

Solving this differential equation gives a nice formula for $\widetilde G$.

\begin{theorem}\label{theo:tildeg}
  The function \(\widetilde{G}(z,u) = \sum_{n \geq 0}\sum_{k = 1}^{n} g_{n,k} 
  \frac{z^n u^k}{k!}\) is equal to
  \[
    \widetilde G (z,u) =
    {\left(1 - \frac{z^2 u}{1 - z}\right)}^{- \frac{1 - z}{z}}.
  \]
\end{theorem}

By a tedious but straightforward application of the transfer theorem~\cite{FS2009}, we can compute the average number of black vertices under the \labelled\  distribution.

\begin{proposition}

  Let $\main{\gamma_n}$ be the number of commits in the main branch of a graph
  $\gamma_n$ taken at random with probability~\(\mathbb P(\gamma_n) =
  \frac{u^{\main{\gamma_n}}}{\main{\gamma_n}!} \frac{1}{\widetilde{G_n}(u)}\).
  The mean and variance of $\main{\gamma_n}$ are asymptotically equivalent to

  \[
    \mathbb E (\main{\gamma_n}) \sim \frac{1 - \rho_u }{2 - \rho_u} n
    \quad \text{ and } \quad
    \mathbb V (\main{\gamma_n}) \sim \frac{\rho_u (1 - \rho_u) }{ {(2 - \rho_u)}^3} n, \quad \text{ where } \rho_u = \frac{\sqrt{1 + 4 u} - 1}{2u}.
  \]
  
\end{proposition}

  Remark that the expected value of the $\main \gamma / n$ ratio can be any number between $0$ and $1/2$, de\-pending on the value of~$u$. 
  This is one of the main benefits of the \labelled\ distribution: given any $\alpha \in (0,\frac 1 2)$, we can \textit{tune}~$u$ in order to target \gitgraph{}s to have $\alpha n$ black vertices (and the variance is quite tight).

\subsection{A bijection with \cyclarium{}s}

The closed formula for $\widetilde G$ featured in Theorem~\ref{theo:tildeg} calls for a combinatorial interpretation. That is why we define a new family of combinatorial objects: the \cyclarium{}s.

\begin{figure}[htb]
    \begin{center}
      \includegraphics[width=0.8\textwidth]{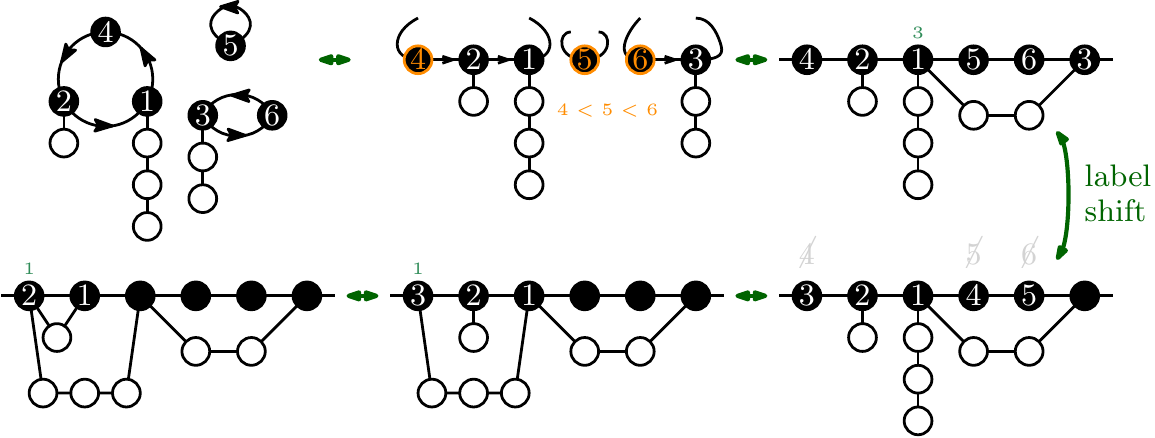}
      \caption{Outline of the bijection between \gitgraph{}s and \cyclarium{}s}%
      \label{fig:bijection}
    \end{center}% 
  \end{figure}

A \emph{\cyclarium} is defined as a set of cycles of $k$ black vertices labeled by $\{1,\dots,k\}$ where each vertex that has not the largest label inside its own cycle carries a non-empty chain of white vertices.
  See Figure~\ref{fig:bijection} top left to see an illustration of a \cyclarium.
  The set $\mathcal Y$ of \cyclarium{}s has the natural combinatorial specification
\begin{equation}\label{eq:spec:cyclariums}
      \mathcal{Y} = \Set{\C}, \quad
      \Seq[\neq 0]{\Z} \times \C = \Cyc{\U\Z \Seq[\neq 0]{\Z}}
 \end{equation}
where $\Set{\cdot}$, $\Seq[\neq 0]{\cdot}$ and $\Cyc \cdot$ are respectively the
operators for sets, non-empty sequences and cycles. Consequently the generating
function of \cyclarium{}s (scaled by $k{!}$) is also given by the formula of
Theorem~\ref{theo:tildeg} (for more details on the symbolic method,
see~\cite{FS2009}).

\begin{proposition}%
  \label{prop:bijection}
  The \gitgraph{}s with $n$ vertices, $k$ black vertices and $f$ free vertices
  are in bijection with the \cyclarium{}s with $n$ vertices, $k$ black vertices
  and $f$ cycles.
\end{proposition}

The bijection is depicted in Figure~\ref{fig:bijection}. 
We give a quick overview of the transformation from \cyclarium{}s to \gitgraph{}s.
First, we break each cycle just before the vertex with the largest label,
 so that they are directed paths. 
 Then we sort these paths according to their largest label,
 and concatenate them. 
 Now we process the black vertices from right to left. 
 If a chain of white vertices is attached to the current black vertex $\boldsymbol v$, 
 then we connect this chain to the black vertex whose position is given by the label of $\boldsymbol v$. 
 If no chain is attached, we do nothing. 
 Once the vertex has been processed, its label $\ell$ is deleted and we change all labels $x$ such that $x > \ell$ by $x-1$. 
 We can check that we eventually obtain a \gitgraph.

Exploiting the fact that the permutations with $f$ cycles are counted by the Stirling numbers of the first kind, we obtain a closed formula for $g_{n,k}$.

  \begin{corollary}%
    \label{cor:formula}
    The number of \gitgraph{}s $g_{n,k}$ of size $n$ and with $k$ black vertices is $1$ if $k = n$ and
    \[g_{n,k} = \sum_{f = 1}^{k-1} {k \brack f} { n - k - 1 \choose k  - f - 1}\]
    for $ k < n$,
    where \({\cdot \brack \cdot}\) denotes the (unsigned) Stirling number of the first kind. 
  \end{corollary}

  \begin{algorithm}[h!]
    \caption{Exact sampler of \gitgraph{}s with $n$ vertices and $k$ black vertices}\label{algo:free}
    \begin{algorithmic}[1]
        \Optional{$f$, the number of free vertices}
        \State{If $f$ is not given, sample it with probability ${k \brack f} { n - k - 1 \choose k  - f - 1}/g_{n,k}$}
        \State{Generate a random permutation of size $k$ with $f$ cycles}
        \State{Generate a composition of $n-k$ into $k-f$ positive terms}
        \State{Form $k-f$ chains of white vertices whose lengths are given by the previous composition}
        \State{Attach them to the permutation to form a \cyclarium}
        \State{Use the bijection from \cyclarium{}s to \gitgraph{}s}
    \end{algorithmic}
  \end{algorithm}

The bijection also suggests a sampling algorithm for \gitgraph{}s of size $n$ if we fix the number $k$ of black vertices and optionally the number $f$ of free vertices: see Algorithm~\ref{algo:free}. 
It runs in $O(n)$ (with some optimization) 
but requires an expensive precomputation of the Stirling numbers of the first kind.
This precomputation is in particular used to generate a uniform permutation of size $k$ with $f$ cycles\footnote{
  The uniform sampler for permutations with a fixed number of cycles is recursive and comes from~\cite[page 33]{Wilf1999} 
  but it might be improved by sampling a Poisson-Dirichlet distribution~\cite[Chapter 3]{Pitman2006} with a well-chosen $\theta$ parameter.
   We leave this as an open question.
  }.
If $f$ must be sampled, we need to precompute $O(k^2)$ numbers of size $O(k \log k)$. If $f$ is given, only $O(f(k-f))$ of them can be precalculated. 
We could also sample $k$ to get a random generator of fixed size $n$.
In this case, the complexity of precomputing the numbers becomes~$O(n^2)$ and those numbers have size~$O(n \log n)$.

\subsection{A Boltzmann generator}

\begin{figure}[htb]
  \begin{center}
    \includegraphics[width=0.99\textwidth]{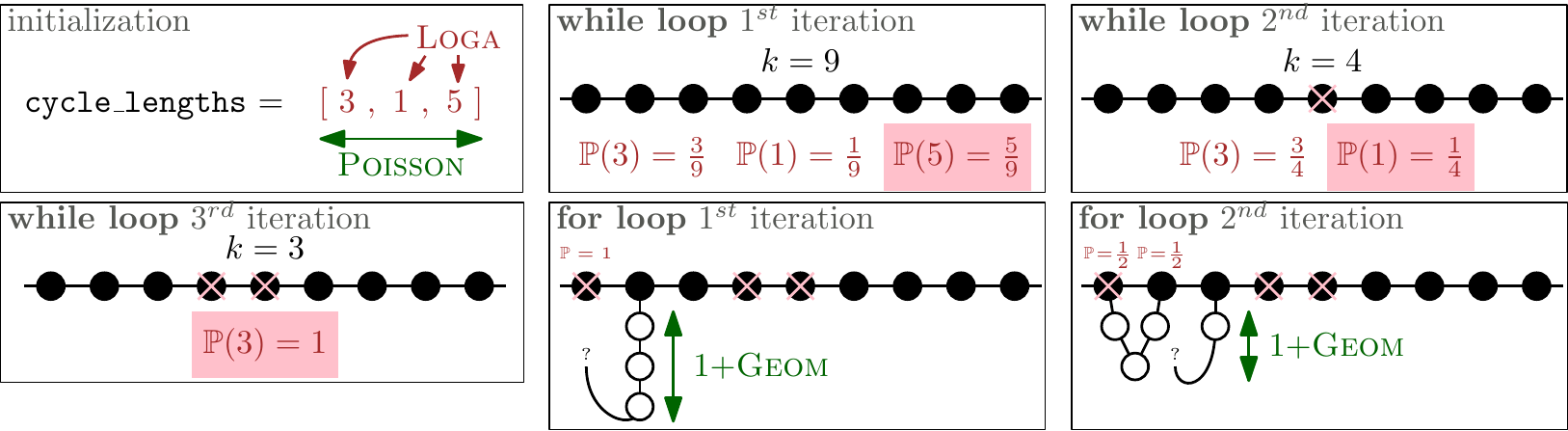}
    \caption{Illustration of the first steps of Algorithm~\ref{algo:boltz2}.}%
    \label{fig:boltzmann}
  \end{center}% 
\end{figure}

Specification~\eqref{eq:spec:cyclariums} induces a natural Boltzmann generator~\cite{DFLS2004} for \cyclarium{}s,
 and hence by Proposition~\ref{prop:bijection} a Boltzmann generator for \gitgraph{}s under the \labelled{} distribution. 
 Rather than simply generating a \cyclarium{} of size $n$ and applying the bijection, which would result in $O(n^2)$ complexity, 
 we can mix the two approaches and achieve $O(n+f^2)$ complexity, where $f$ is
 the number of free vertices (which is logarithmic in~$n$ in average).
 The details are given in Algorithm~\ref{algo:boltz2} and illustrated by Figure~\ref{fig:boltzmann}.

\begin{algorithm}[htb]

  \caption{Boltzmann sampler under the \labelled{} distribution of parameters $z$ and $u$}\label{algo:boltz2}
  \begin{algorithmic}[1]
%    \Require{$0 < u$ and~$0 < z < \rho(u)$}
    % \Function{$\Gamma[\Gt]$}{$z, u$}
      \State{$f \gets \text{\Call{Poisson}{$\ln \Gtf(z, u)$}}$}%
      \Comment{Poisson distribution}
      \State{\texttt{cycle\_lengths} $\gets$ array of~$f$
      independent~\Call{Loga}{$\frac{uz^2}{1 - z}$}}
      \Comment{Logarithmic series distribution}
      \State{$k \gets$ total sum of \texttt{cycle\_lengths}}
      \State{$g \gets$ directed path of $k$ black vertices denoted $v[0],\dots,v[{k-1}]$} \Comment{skeleton of our \gitgraph}
      \While{ $k > 0$ }
        \State{ extract a number $x$ from \texttt{cycle\_lengths} with probability $x/k$}
        \State{ mark $v[{k-x}]$  }
        \State{ $k \gets k - x$}
      \EndWhile{}

      \For{$j$ \textbf{from}~$1$ \textbf{to}~number of black vertices $-1$}
        \If{$v[j]$ is \textbf{not} marked}%
          \State{$i \gets$ random number between $0$ and $j-1$}             
          \State{link $v[i]$ to a directed path of $(1 + \text{\Call{Geom}{$z$}})$ white vertices}\Comment{Geometric distribution}
          \State{link the last vertex of this path to $v[j]$}
        \EndIf{}
      \EndFor{}
      \State\Return{$g$}
    % \EndFunction{}
  \end{algorithmic}
\end{algorithm}

Our implementation of this algorithm in Python easily generates graphs larger than 10 million. 
We also recall that we can carefully choose the parameters $z$ and $u$ to target a size $n$ and a ratio $\alpha \in (0,\frac 1 2)$, where $\alpha n$ is the number of black vertices.

\section{Conclusion}

In this work, we have developed three random generators for \gitgraph{}s.

A few questions remain unanswered.
Firstly, our algorithms are unable to generate graphs for certain values of $k$ efficiently (more precisely when $k$ is in the window $\sqrt n \ll k \ll n$, and when $k \geq \frac n 2$). 
In addition, it would be interesting to obtain an asymptotic estimate of the numbers $g_n$ of \gitgraph{}s.
The formula in Corollary~\ref{cor:formula} seems to be a good start to do so.
Moreover, we could study potential phase transitions as $k$ evolves as a function of $n$. 
In particular, we could investigate how the number of free vertices grows, 
as well as the gaps between each of them.

Finally, we could study more involved workflows, and enumerate DAGs that adhere to them.

\bibliography{biblio}{}
\bibliographystyle{eptcs}

\end{document}

%% file: preamble.tex
\usepackage{graphicx}

\usepackage{xcolor}
\definecolor{Carmine}{HTML}{9D000C}
\definecolor{Pakistan Green}{HTML}{0F6B00}
\definecolor{Sapphire Blue}{HTML}{006BA6}
\definecolor{darkorange}{rgb}{1,0.549,0}

\usepackage{amsmath}
\usepackage{amsthm}
\usepackage{amssymb}
\usepackage{mathtools}

\newtheorem{corollary}{Corollary}
\newtheorem{definition}{Definition}

\newtheorem{theorem}{Theorem}
\newtheorem{proposition}{Proposition}

\usepackage{algorithm}
\usepackage[noend]{algpseudocode}

\algnewcommand\algorithmicinput{\textbf{Input:}}
\algnewcommand\algorithmicoutput{\textbf{Output:}}
\algnewcommand\Input{\item[\algorithmicinput]}
\algnewcommand\Output{\item[\algorithmicoutput]}
\algnewcommand\algorithmicOptional{\textbf{Additional optional input:}}
\algnewcommand\Optional{\item[\algorithmicOptional]}
\usepackage{cite}

\newcommand\C{\mathcal{C}}

\newcommand\U{\mathcal{U}}
\newcommand\Z{\mathcal{Z}}

\newcommand\MetaOp[3]{\ensuremath{\textsc{#1}_{#2}\left(#3\right)}}
\newcommand\Cyc[2][]{\MetaOp{Cyc}{#1}{#2}}
\newcommand\Seq[2][]{\MetaOp{Seq}{#1}{#2}}
\newcommand\Set[2][]{\MetaOp{Set}{#1}{#2}}
\newcommand\Gtf{\tilde{G}}

\newcommand\gitgraph{Git graph}
\newcommand\fullnamegitgraph{Git feature branch graph}
\newcommand\main[1]{\ensuremath{k(#1)}}
\newcommand\cyclarium{cyclarium}
\newcommand\labelled{labeled-main}

%% file: main.bbl
\begin{thebibliography}{1}
\providecommand{\bibitemdeclare}[2]{}
\providecommand{\surnamestart}{}
\providecommand{\surnameend}{}
\providecommand{\urlprefix}{Available at }
\providecommand{\url}[1]{\texttt{#1}}
\providecommand{\href}[2]{\texttt{#2}}
\providecommand{\urlalt}[2]{\href{#1}{#2}}
\providecommand{\doi}[1]{doi:\urlalt{https://doi.org/#1}{#1}}
\providecommand{\eprint}[1]{arXiv:\urlalt{https://arxiv.org/abs/#1}{#1}}
\providecommand{\bibinfo}[2]{#2}

\bibitemdeclare{article}{Borel1899}
\bibitem{Borel1899}
\bibinfo{author}{{\'E}mile \surnamestart Borel\surnameend}
  (\bibinfo{year}{1899}): \emph{\bibinfo{title}{M{\'e}moire sur les s{\'e}ries
  divergentes}}.
\newblock {\slshape \bibinfo{journal}{Annales scientifiques de l'{\'E}cole
  Normale Sup{\'e}rieure}} \bibinfo{volume}{3e s{\'e}rie, 16}, pp.
  \bibinfo{pages}{9--131}, \doi{10.24033/asens.463}.

\bibitemdeclare{article}{CDL2023}
\bibitem{CDL2023}
\bibinfo{author}{Julien \surnamestart Courtiel\surnameend},
  \bibinfo{author}{Paul \surnamestart Dorbec\surnameend} \&
  \bibinfo{author}{Romain \surnamestart Lecoq\surnameend}
  (\bibinfo{year}{2023}): \emph{\bibinfo{title}{Theoretical Analysis of Git
  Bisect}}.
\newblock {\slshape \bibinfo{journal}{Algorithmica}},
  \doi{10.1007/s00453-023-01194-0}.

\bibitemdeclare{article}{DFLS2004}
\bibitem{DFLS2004}
\bibinfo{author}{Philippe \surnamestart Duchon\surnameend},
  \bibinfo{author}{Philippe \surnamestart Flajolet\surnameend},
  \bibinfo{author}{Guy \surnamestart Louchard\surnameend} \&
  \bibinfo{author}{Gilles \surnamestart Schaeffer\surnameend}
  (\bibinfo{year}{2004}): \emph{\bibinfo{title}{Boltzmann samplers for the
  random generation of combinatorial structures}}.
\newblock {\slshape \bibinfo{journal}{Combinatorics, Probability \& Computing}}
  \bibinfo{volume}{13}(\bibinfo{number}{4-5}), pp. \bibinfo{pages}{577--625},
  \doi{10.1017/S0963548304006315}.

\bibitemdeclare{book}{FS2009}
\bibitem{FS2009}
\bibinfo{author}{Philippe \surnamestart Flajolet\surnameend} \&
  \bibinfo{author}{Robert \surnamestart Sedgewick\surnameend}
  (\bibinfo{year}{2009}): \emph{\bibinfo{title}{Analytic Combinatorics}}.
\newblock \bibinfo{publisher}{Cambridge University Press},
  \doi{10.1017/CBO9780511801655}.
\newblock \urlprefix\url{https://algo.inria.fr/flajolet/Publications/book.pdf}.

\bibitemdeclare{article}{FZV1994}
\bibitem{FZV1994}
\bibinfo{author}{Philippe \surnamestart Flajolet\surnameend},
  \bibinfo{author}{Paul \surnamestart Zimmermann\surnameend} \&
  \bibinfo{author}{Bernard \surnamestart {Van Cutsem}\surnameend}
  (\bibinfo{year}{1994}): \emph{\bibinfo{title}{A calculus for the random
  generation of labelled combinatorial structures}}.
\newblock {\slshape \bibinfo{journal}{Theoretical Computer Science}}
  \bibinfo{volume}{132}(\bibinfo{number}{1-2}), pp. \bibinfo{pages}{1--35},
  \doi{10.1016/0304-3975(94)90226-7}.

\bibitemdeclare{phdthesis}{LecoqThesis}
\bibitem{LecoqThesis}
\bibinfo{author}{Romain \surnamestart Lecoq\surnameend} (\bibinfo{year}{2024}):
  \emph{\bibinfo{title}{Le feu {\c c}a br{\^u}le et l'informatique {\c c}a
  bugue : Combustion et r{\'e}gression dans les graphes}}.
\newblock Ph.D. thesis.
\newblock \bibinfo{note}{In French, work in progress}.

\bibitemdeclare{book}{NW1978}
\bibitem{NW1978}
\bibinfo{author}{Albert \surnamestart Nijenhuis\surnameend} \&
  \bibinfo{author}{Herbert \surnamestart Wilf\surnameend}
  (\bibinfo{year}{1978}): \emph{\bibinfo{title}{Combinatorial Algorithms: For
  Computers and Hard Calculators}}, \bibinfo{edition}{2nd} edition.
\newblock \bibinfo{publisher}{Academic Press, Inc.}, \bibinfo{address}{USA},
  \doi{10.1016/C2013-0-11243-3}.

\bibitemdeclare{book}{Pitman2006}
\bibitem{Pitman2006}
\bibinfo{author}{Jim \surnamestart Pitman\surnameend} (\bibinfo{year}{2006}):
  \emph{\bibinfo{title}{Combinatorial stochastic processes}}.
\newblock {\slshape \bibinfo{series}{Lecture Notes in Mathematics}}
  \bibinfo{volume}{1875}, \bibinfo{publisher}{Springer-Verlag, Berlin},
  \doi{10.1007/b11601500}.
\newblock \bibinfo{note}{Lectures from the 32nd Summer School on Probability
  Theory held in Saint-Flour, July 7--24, 2002, With a foreword by Jean
  Picard}.

\bibitemdeclare{unpublished}{Wilf1999}
\bibitem{Wilf1999}
\bibinfo{author}{Herbert~S. \surnamestart Wilf\surnameend}
  (\bibinfo{year}{1999}): \emph{\bibinfo{title}{East Side, West Side . . . - an
  introduction to combinatorial families-with Maple programming}}.
\newblock \urlprefix\url{https://www2.math.upenn.edu/~wilf/eastwest.pdf}.

\end{thebibliography}
